\documentclass[aps,prd,reprint,nofootinbib,longbibliography]{revtex4-1}
\usepackage{blindtext}
\usepackage{mathtools}
\usepackage{xcolor}
\usepackage{adjustbox}
\usepackage{graphicx}

\usepackage[utf8]{inputenc}
\usepackage[english]{babel}
\usepackage{hyperref}
\hypersetup{
    colorlinks=true,
    linkcolor=blue,
    citecolor=red,
    filecolor=magenta,      
    urlcolor=magenta,
}

\begin{document}
\title{Can QCD Axion Stars explain Subaru HSC microlensing?}
\author{Enrico D. Schiappacasse$^{1,2\dagger}$ and Tsutomu T. Yanagida$^{3\ddagger}$}

\affiliation{
$^1$ Department of Physics, University of Jyv$\ddot{a}$skyl$\ddot{a}$, P.O.Box 35 (YFL), FIN-40014 Jyv$\ddot{a}$skyl$\ddot{a}$,  Finland\\
$^2$Helsinki Institute of Physics, University of Helsinki, P.O. Box 64, FIN-00014 Helsinki, Finland\\ 
$^3$Tsung-Dao Lee Institute and School of Physics and Astronomy, Shanghai Jiao Tong University, 
 200240 Shanghai, China }

\begin{abstract}
A non-negligible fraction of the QCD axion dark matter may form   
gravitationally bound Bose Einstein condensates, which are commonly known as axion stars or axion clumps. Such 
astrophysical objects have been recently proposed as the cause for the single candidate event reported by Subaru Hyper Suprime-Cam (HSC) microlensing search in the Andromeda galaxy. Depending on the breaking scale of the Peccei–Quinn symmetry and the details of the dark matter scenario, QCD axion clumps may form via gravitational condensation during radiation domination, in the dense core of axion miniclusters, or within axion minihalos around primordial black holes. We analyze all these scenarios and conclude that the microlensing candidate detected by the Subaru HSC survey is likely not caused
by QCD axion stars.
\end{abstract}

\maketitle

\section{Introduction}
The current leading particle dark matter (DM) candidate is the QCD axion~\cite{PRESKILL1983127, 1983PhLB..120..133A, DINE1983137}, being strongly motivated by shortcomings in the standard model of particle physics~\cite{PhysRevLett.38.1440, PhysRevLett.40.223, PhysRevLett.40.279} and unification ideas in the frame of string theory~\cite{Svrcek:2006yi, Douglas:2006es, Arvanitaki:2009fg}. Technically speaking, the QCD axion is a gauge singlet (pseudo)scalar coming from the spontaneous breaking of the U(1)-global Peccei-Quinn (PQ) symmetry introduced to solve the strong \textit{CP} problem. The fact that only a small part of the most obvious regime of the axion parameter space has been excluded via the axion DM Experiment~\cite{PhysRevLett.120.151301} encourages one to look for signals in different contexts, including  astrophysics. 

A major role in such setups is played by the so-called axion clumps or axion stars (we will use both terms interchangeably in this paper), which correspond to nonrelativistic Bose Einstein condensates (BECs) of axions. Depending on the energy scale at which the PQ symmetry is broken, e.g., before or after inflation,  these clumps may form via gravitational condensation after thermalization in the early Universe~\cite{Guth:2014hsa, Schiappacasse:2017ham, Visinelli:2017ooc}, nucleation in the core of axion miniclusters~\cite{Schive:2014hza} or nucleation in the inner shells of minihalos around primordial black holes (PBHs)~\cite{Hertzberg:2020hsz}~\footnote{ The study of PBHs
dates back to the 1960s and it was soon realised~\cite{1975Natur.253..251C} that they are a good DM candidate with a profuse
phenomenology associated with them~\cite{Kawasaki:1997ju, Carr:2016drx, Eroshenko:2016yve, Hertzberg:2020kpm, Nurmi:2021xds, Kainulainen:2021rbg}.}. A rich phenomenology is associated with these compact objects ranging from parametric resonance of photons~\cite{Hertzberg:2018zte, Hertzberg:2020dbk, Amin:2021tnq} and explosion in relativistic axions~\cite{Levkov:2016rkk, Eby:2021ece} to axion-photon conversion in neutron star magnetospheres~\cite{Bai:2021nrs}. In addition to   
these indirect searches, QCD axion stars could potentially be discovered by gravitational microlensing events as has been recently suggested in the literature~\cite{Sugiyama:2021xqg}. 

Generally speaking, a gravitational microlensing event refers to the brightness amplification of a background star when a compact object passes close to the line of sight to that star~\cite{1986ApJ...304....1P}. Microlensing from pointlike objects is simple of calculating by using the standard equations for gravitational lensing~\cite{1991ApJ...366..412G}. When the observer, lens, and source
lie on the same line, the radius of closest approach of photons to the point mass lens as these photons pass by it defines the so-called Einstein radius $R_E$. The Einstein radius associated with a point mass $M$ reads as 
\begin{align}
  R_E &= \sqrt{\frac{4 G_N M}{c^2}\frac{D_L D_{\text{LS}}}{D_S}}\,,\\ 
      &=4.3\times 10^{8}\,\text{km}\left( \frac{M}{M_{\odot}}\right)^{1/2}\left( \frac{D_S}{\text{kpc}} \right)^{1/2}\,,~\label{ER}
\end{align}
where we have taken $D_{\text{L}}\sim D_{\text{LS}} \sim D_{\text{S}}$. Here $D_{\text{S}}$, $D_{\text{L}}$, and $D_{\text{LS}}$ are the distances between the source and the observer, the lens and the observer, and the source and the lens, respectively.

Niikura and collaborators~\cite{Niikura:2017zjd} 
report a single candidate for microlensing of stars ($D_\text{s}\sim 770\, \text{kpc}$) in the Andromeda galaxy (M31) after a dense-cadence, 7-hr-long observation of M31 with the Subaru Hyper Suprime-Cam (HSC). They propose primordial black holes in the M31 and Milky Way (MW) galactic halos as the cause for such an event constraining the PBH abundance in the mass range  $\sim[10^{-10}-\textcolor{black}{10^{-5}}]M_{\odot}$ and excluding a PBH fraction of dark matter $f_{\text{PBH}}\gtrsim \mathcal{O}(\text{few})\times 0.01$~\cite{Carr:2020gox}. Here we point out that these constraints and exclusion hold for any compact object which behaves as a pointlike object from the microlensing perspective (under the assumption that such a object follows the spatial and velocity distributions predicted by the standard DM halo model for the MW and M31). 

Assuming QCD axion stars behave as pointlike lens, authors in  Ref.~\cite{Sugiyama:2021xqg} claim that such astrophysical objects could explain the single microlensing candidate reported by the Subaru HSC. The corresponding contour for the credible region at the $95\%$ of confidence includes axion stars with masses around \textcolor{black}{$[4\times10^{-10}-10^{-5}M_{\odot}]$} representing a fraction of the dark matter \textcolor{black}{$f_{\star} \gtrsim 4\times 10^{-6}$}. 

In this article, we study such a claim analyzing the  different known formation mechanisms of QCD axion stars. We conclude that these compact objects, probably, would not cause \textcolor{black}{the microlensing candidate detected by the Subaru HSC observation}. 

\section{Postinflationary PQ symmetry-breaking Scenario}

\subsection{Gravitational condensation in the early Universe}

If the PQ symmetry is broken after inflation, large amplitude axion field fluctuations and topological defects are found on size scales of order the horizon at the time of the  spontaneous symmetry breaking (SSB)~\cite{PhysRevLett.48.1156}. Because of causality reasons, the axion field keeps uncorrelated from one Hubble patch to the next. The axion is massless at the classical level, but it acquires a mass at the time of the QCD phase transition due to instanton effects~\cite{RevModPhys.53.43}. From this point, the large axion field fluctuations allow the field to undergo strong mode-mode interactions via gravity with a relaxation rate $\Gamma_{\text{cond}}\sim 8\pi G_N \textcolor{black}{m_{a}^2 n_{a}}/k^2$, where \textcolor{black}{$m_{a}$ and $n_{a}$} are the axion mass and number density, respectively, $G_N$ is the Newton gravitational constant as usual, and $k$ is a characteristic wave number. The physical wave number evolves under standard redshifting until the relaxation rate becomes comparable to the Hubble rate, e.g., $\Gamma_{\text{cond}}\sim H$. At such a point, the axion system will tend to thermalize. Since cosmological axions are produced at high occupancy and are
nonrelativistic, the temperature of the system is well below the critical temperature for Bose Einstein condensation~\cite{Guth:2014hsa}, so that the system will reorganize itself into a type of BEC after thermalization. 

The true BEC corresponds to spherically symmetric axion configurations resulting from the equilibrium between the attractive gravitational and self-interacting forces with the gradient pressure. The full space of solutions in the parameter space constituted by the axion star mass and radius ($M_{\star}$,$R_{\star}$) is given by two nonrelativistic branches and one relativistic branch, as shown Figs.~1 and 7 in Ref.~\cite{Schiappacasse:2017ham}. Long-lived configurations are stable under radial perturbations and are well described by a nonrelativistic approximation for the axion field. The gravitational attraction is dominant over the self-interaction and the smaller the clump size, the denser the clump. There is a maximum clump mass associated with a minimum clump size both depending on the axion mass and the symmetry-breaking scale. The other nonrelativistic branch has spatially smaller clumps which are unstable under radial perturbations. For extremely smaller clumps, this later branch connects with a relativistic (dense) branch known as axitons~\cite{Kolb:1993hw}, which is quasistable by the profuse emission of relativistic axions. 
\begin{figure}[t!]
\centering
\includegraphics[width=7cm,height=5 cm]{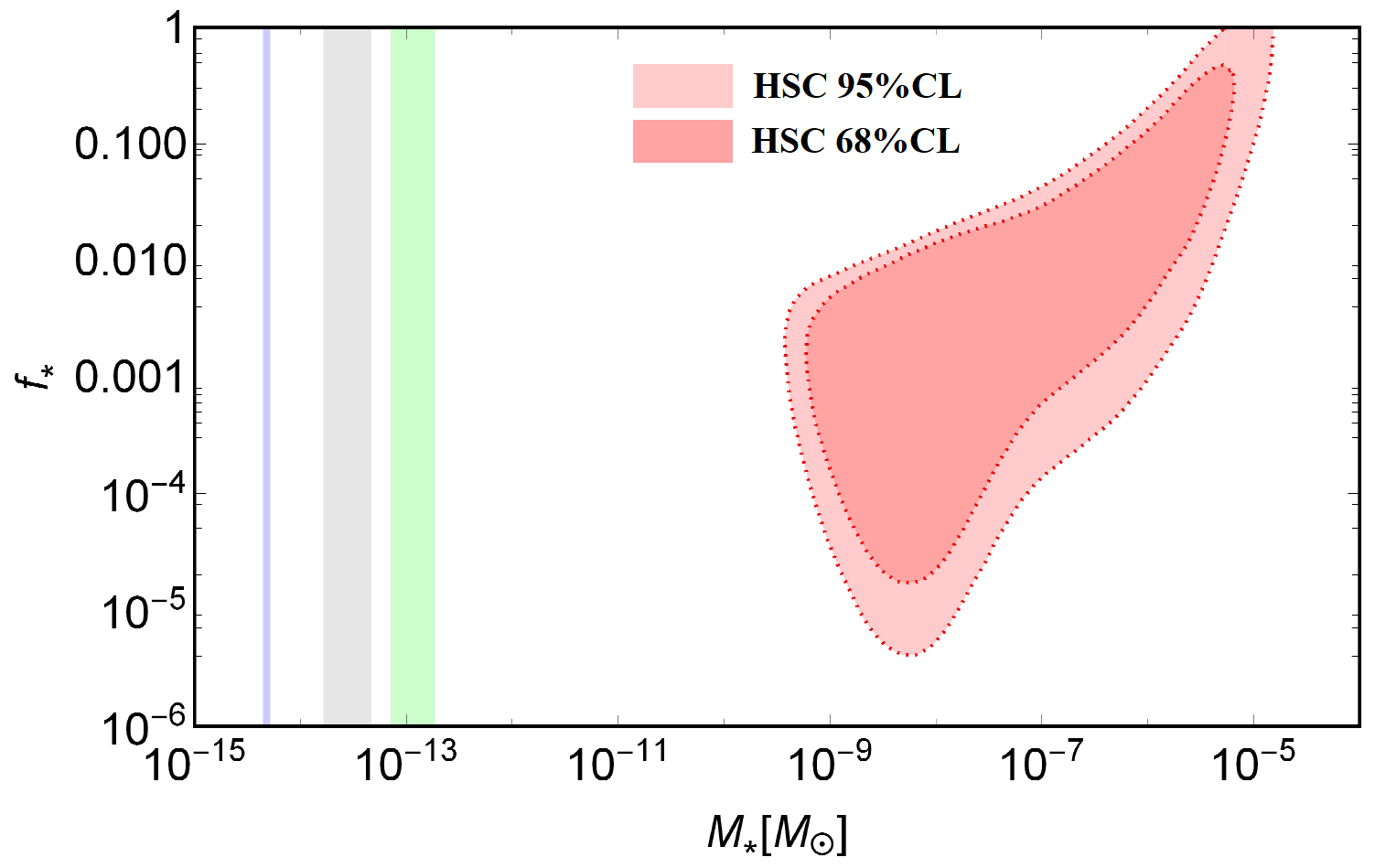}
\caption{In the parameter space $(M_{\star},f_{\star})$, where $f_{\star}$ is the fraction of dark matter in axion stars, the red and pink shaded zones are the $68\%$ and $95\%$ credible regions, respectively, which are consistent with the microlensing candidate reported by the Subaru HSC  observation~\cite{Niikura:2017zjd}, as shown Fig. 1 in Ref.~\cite{Sugiyama:2021xqg}. We have added the mass regime for the maximum axion star mass according to the allowed \textcolor{black}{mass range for the dark matter QCD axion}. The green band uses the axion mass range obtained in numerical simulations from Ref.~\cite{Kawasaki:2014sqa}. The blue and gray \textcolor{black}{bands} use results from Ref.~\cite{Hoof:2021jft}, where a statistical interference of the axion mass window is obtained using the data from the GHV and HKSSYY groups, respectively (see main text). All the above cases assume $N_{\text{DW}}=1$.}  
\label{Q1}
\end{figure}
From the microlensing point of view, the long-lived nonrelativistic axion clump branch is of the most importance and we focus on it from now on.

The QCD axion clumps \textcolor{black}{are ground state solutions of the Schr$\ddot{\text{o}}$dinger-Poisson equations. Using a localized ansatz for the axion radial profile depending on one single parameter, the scale length $R$, and a variational approach by extremizing the nonrelativistic Hamiltonian at fixed number of particles, the physical mass and radius of (long-lived) stationary solutions} are given, respectively, by~\cite{Fujikura:2021omw}~\footnote{\textcolor{black}{Here we update the physical parameters obtained in Ref.~\cite{Hertzberg:2020dbk},  Eqs.~(2.23)-(2.27), by using an exponential-linear ansatz for the axion clump radial profile, which is slightly more accurate than the sech ansatz profile previously used \textcolor{black}{(see Ref.~\cite{2011PhRvD..84d3531C} for an earlier estimate in a self-gravitating BEC with
short-range interactions)}.}}
\begin{equation}
\begin{adjustbox}{max width=223pt}
 $M_{\star} \approx 1.2\times 10^{-11}\,M_{\odot} \times \alpha \left( \frac{10^{-5}\,\text{eV}}{m_{a}} \right) \left( \frac{F_a}{6\times10^{11}\,\text{GeV}} \right) \left( \frac{0.3}{\gamma} \right)^{1/2}\,,$\label{Mmax}  
 \end{adjustbox}
\end{equation}
\begin{equation}
\begin{adjustbox}{max width=223pt}
$R^{90}_{\star} \approx 78\,\text{km}\left( \frac{1+\sqrt{1-\alpha^2}}{\alpha} \right) \left( \frac{10^{-5}\,\text{eV}}{m_{a}} \right) \left( \frac{6\times10^{11}\,\text{GeV}}{F_a} \right)  \left( \frac{\gamma}{0.3} \right)^{1/2}\,, $~\label{rstar} 
\end{adjustbox}
\end{equation}
where the QCD axion mass is defined as $m_a = 6\,\mu \text{eV} (10^{12}\text{GeV}/F_a)$~\cite{PhysRevLett.40.223} and $\gamma = 1 - 3m_um_d/(m_u+m_d)^2\approx 0.3$ comes from the expansion of the QCD potential around the \textit{CP} preserving vacuum~\cite{GrillidiCortona:2015jxo}, with $m_u$ and $m_d$ as the up and down quark masses, respectively. Since axion clumps do not have a hard surface, we have taken
$R^{90}_{\star}$ as the radius which encloses $0.9M_{\star}$. The parameter $\alpha \in\, ]0,1]$ allows us
to go over the branch so that $M_{\star,\text{max}}=M_{\star}(\alpha=1)$, where $M_{\star,\text{max}}$ is the maximum allowed mass for an axion clump. 

In this postinflationary scenario for the PQ symmetry, there is an additional axion abundance coming from the decay of topological defects, which significantly modify the usual axion abundance coming from the misalignment mechanism. Numerical simulations performed in Ref.~\cite{Kawasaki:2014sqa} show that the QCD axion can be responsible for the cold dark matter in the Universe in the mass range $m_{a} = [0.8-1.3]\times10^{-4}\text{eV}$ when the domain wall number $(N_{\text{DW}})$ is equal to unity. When $N_{\text{DW}}>1$ and a mild tuning of the model parameters are allowed, we have $m_{a} = \mathcal{O}(10^{-4}-10^{-2})\text{eV}$.  Recently, a statistical analysis of different results from QCD axion string and domain walls simulations was performed in Ref.~\cite{Hoof:2021jft} for the particular case of $N_{\text{DW}}=1$. \textcolor{black}{For DM axions}, the $95\%$ credible interval at highest posterior density using the data from Refs.~\cite{Gorghetto:2020qws, Gorghetto:2018myk} (GHV group of results) and Refs.~\cite{Hiramatsu:2012gg, Hiramatsu:2010yu, Kawasaki:2014sqa} (HKSSYY group of results) is $m_{a}=[4.8-5.2]\times10^{-4}\text{eV}$ and $m_{a}=[1.6-2.7]\times10^{-4}\text{eV}$, respectively. 

Let $f_{\star}$ be the fraction of dark matter in axion stars. In the parameter space $(M_{\star},f_{\star})$, Fig.~1 shows the $68\%$ (red shaded region) and $95\%$ (pink shaded region) credible regions obtained in Ref.~\cite{Sugiyama:2021xqg} to explain the single microlensing event reported by Subaru HSC observation. Taking $\alpha=1$ in Eq.~(\ref{Mmax}), we have calculated the maximum mass which can be reached by a QCD axion star considering the different axion mass windows discussed above and shown in Fig. 1 as colored green, blue, and gray bands. We see that $M_{\star,\text{max}} \lesssim 10^{-13}M_{\odot}$, which is about 3 orders of magnitude below the lighter masses required to explain the Subaru HSC observation~\footnote{Here we mention that for the case of axionlike particle stars the axion mass and symmetry-breaking scale are unrelated so that both can be taken as free parameters. Then, the axion star mass may reach values within the credible regions to explain HSC data. }.

Here we point out that, after thermalization, the axion field may condense in higher eigenstates with nonzero angular momentum. These axion rotating stars hold the relation $l=|m|$ for their spherical harmonic numbers, since this state minimizes the energy of the system at fixed angular momentum and number of particles. For large angular momentum, the maximum mass allowed for stable rotating configurations increases rapidly  as $M_{\star,\text{max}} \propto l^{3/2}/(\text{ln} l)^{1/4}$, but the associated minimum radius increases slowly $R_{\star,\text{min}} \propto l^{1/2}/(\text{ln} l)^{1/4}$~\cite{Hertzberg:2018lmt}. For angular momentum $l=|m|>\mathcal{O}(10^2)$, we will have $M_{\star,\text{max}} > \mathcal{O}(10^{-10})M_{\odot}$, e.g., within the credible region to explain Subaru HSC data. However, given the fact that this is a large amount of angular momentum and numerical simulations have shown that these configurations tend to release the excess of angular momentum by particles ejection~\cite{Hertzberg:2020dbk, Dmitriev:2021utv}, such a situation seems improbable. 

\subsection{Nucleation in QCD axion miniclusters}

The cosmology and astrophysics of QCD axion miniclusters were explored in detail in the early 1990s~\cite{Hogan:1988mp, Kolb:1993zz, Kolb:1993hw, Kolb:1994fi, Kolb:1995bu}. The large axion field fluctuations \textcolor{black}{produced} by the SSB remain smooth up to scales of order the horizon size at the time when the axion acquires its mass. From this point, density perturbations may grow under gravity as usual to collapse into gravitationally bound DM substructures known as miniclusters at around matter-radiation equality~\cite{Zurek:2006sy}. Numerical simulations have shown that a sizable fraction of the axion as DM may end up in such bound structures~\cite{Vaquero:2018tib, Buschmann:2019icd}.  Axion miniclusters are not able to form in the preinflationary scenario even though if the QCD axion initial conditions are extremely fine-tuned~\cite{Fukunaga:2020mvq}. 

Because of the randomness of the initial axion overdensity fluctuations and the subsequent evolution of the minicluster halo mass function via hierarchical structure formation during matter domination, masses of axion miniclusters span several orders of magnitude as $10^{-19}M_{\odot}\lesssim M_{\text{AMC}}\lesssim 10^{-5}M_{\odot}$~\cite{Kavanagh:2020gcy}.

\textcolor{black}{Neglecting the axion self-interaction, numerical simulations performed by Refs.~\cite{Levkov:2018kau, PhysRevD.100.063528} have shown that} axion clumps may kinetically nucleate in the dense central region of axion miniclusters via gravitational interactions, when the axion field coherence length is much smaller than the minicluster radius. \textcolor{black}{Such axion stars} hold similar radial profiles and large-amplitude oscillations to solitonic cores found in simulations of fuzzy dark matter halos~\cite{Schive:2014hza, PhysRevD.98.043509}. The solitonic core mass and the mass of its host halo are related to each other by the expression~\cite{Schive:2014hza}
\textcolor{black}{\begin{align}
M_{\star} = &1.4 \times 10^{-16}M_{\odot}(1+z_{\star})^{1/2}\nonumber\\
&\times\left( \frac{\zeta(z_{\star})}{\zeta(0)}  \right)^{1/6}\left(\frac{10^{-5}\text{eV}}{m_a} \right) \left( \frac{M_{\text{AMC}}}{10^{-10}\,M_{\odot}} \right)^{1/3}\,,\label{ms} 
\end{align}}
\textcolor{black}{\hspace{-0.1 cm}where $z_{\star}$ is the redshift at the formation time, $\zeta(z_{\star}) \simeq 18\pi^2$ for $z_{\star}\gg1$ and $\zeta(0)\sim350$~\cite{Schive:2014hza, Bryan:1997dn}, with $\zeta(z)\equiv (18\pi^2+82(\Omega_m(z)-1)-39(\Omega_m(z)-1)^2)/\Omega_m(z)$.} 
The mass and radius of the soliton core are inversely proportional as~\cite{Schive:2014hza, PhysRevD.100.063528}
\begin{equation}
 R_{\star} =  1.5 \,\text{km} \left( \frac{10^{-10}M_{\odot}}{M_{\star}}\right) \left(\frac{10^{-5}\text{eV}}{m_{a}} \right)^2 \,.\label{m-r} 
\end{equation}

Numerical simulations performed in Ref.~\cite{PhysRevD.100.063528} report nucleation events in the center of miniclusters \textcolor{black}{at $z_{\star}\sim (1280-900)$}, including a double nucleation in one minicluster with two maxima. 
The nucleated axion stars end up surrounded by density waves in the incoherent granular density fluctuations of miniclusters. \textit{A priori}, there is no reason to consider that axion stars will leave the minicluster environment after nucleation so that they may act as isolated lensing objects. Even if this is the case, we see from Eq.~(\ref{ms}) that, for typical minicluster masses \textcolor{black}{and nucleation time}, the QCD axion star masses would be \textcolor{black}{a few} orders of magnitude below than the lighter masses needed to explain HSC observation.

Lastly, we point out that axion star masses within the credible region to explain HSC data in Fig.~1 are much heavier than
the maximum mass allowed for an axion star, Eq.~(\ref{Mmax}) with $\alpha = 1$. Axion stars
with $M_{\star}> M_{\star,\text{max}}$ are unstable under collapse partially exploding in relativistic axions to return later to the non-relativistic stable branch with $M_{\star}\lesssim M_{\star,\text{max}}$ ~\cite{2017PhRvL.118a1301L}.

\section{Preinflationary PQ symmetry-breaking Scenario}

When the PQ symmetry is broken before (or during) inflation, the axion field is driven to be highly homogeneous on large scales. In such a scenario, axion clumps may still form if there is a small fraction of DM in PBHs, as shown in Ref.~\cite{Hertzberg:2020hsz}. 

Even though PBHs have not been proved yet, their existence has been suggested by the gravitational wave events detected by LIGO-Virgo Collaboration~\cite{LIGOScientific:2017bnn, LIGOScientific:2017zid, LIGOScientific:2016ebi, LIGOScientific:2016vpg, LIGOScientific:2016aoc}, the recent NANO-Grav results~\cite{NANOGrav:2020bcs, Kohri:2020qqd, Vaskonen:2020lbd, Inomata:2020xad}, and fast radio bursts~\cite{Kainulainen:2021rbg}. In a mixed DM scenario composed of a dominant axion plus a small fraction of DM in PBHs, these compact objects will unavoidably acquire axion minihalos based on the secondary infall accretion mechanism~\cite{1985ApJS...58...39B, Mack:2006gz}. \textcolor{black}{While the virialized minihalo mass $M_{\text{halo}}$ and radius $R_{\text{halo}}$ grow with the redshift as $M_{\text{halo}}(z) \propto (1+z)^{-1}M_{\text{PBH}}$ and $R_{\text{halo}}\propto(1+z)^{-4/3}M_{\text{PBH}}^{1/3}$, respectively, where $M_{\text{PBH}}$ is the mass of the central PBH~\cite{Mack:2006gz, Ricotti:2007au}, the self-similar minihalo profile reads as $\rho_{\text{halo}}\sim \mathcal{O}(10^{-1})M_{\odot}\text{pc}^{-3}(R_{\text{halo}}/r)^{9/4}(10^2 M_{\text{PBH}}/M_{\text{halo}})^3$~\cite{Nurmi:2021xds}.}
 
Such minihalos satisfy the required conditions to kinetically form axion clumps before the time of first galaxies formation. The relaxation rate for such a nucleation reads as $\Gamma_{\text{kin}} \sim n_{a}\sigma_{\text{gr}}v_{a}\mathcal{N}$~\cite{Levkov:2018kau} with $\sigma_{\text{gr}}\propto (G_N m_{a}/v_{a}^2)^2$, where $v_{a}$ and $\mathcal{N}$ are the axion virial velocity in minihalos and the occupancy number associated with the Bose enhancement, respectively, and $\sigma_{\text{gr}}$ is the gravitational scattering cross section. 
\begin{figure}[t!]
\centering
\includegraphics[width=7.5cm,height=5.5 cm]{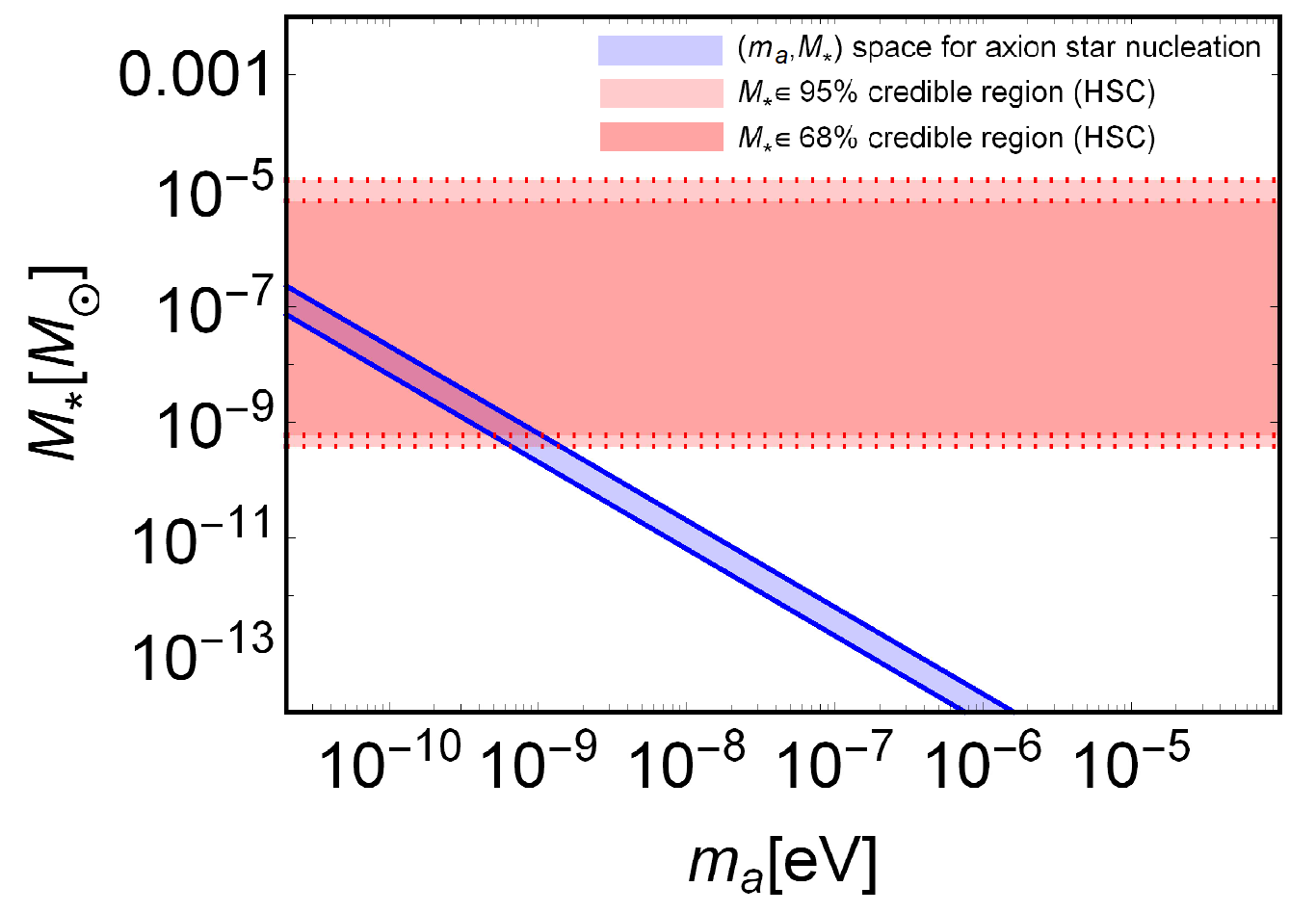}
\caption{The blue shaded region corresponds to the parameter space $(m_{a},M_{\star})$ at which axion minihalos around PBHs may kinetically nucleate axion stars before the formation of first galaxies (more details are found in Ref.~\cite{Hertzberg:2020hsz}, Sec. 3). As additional information, we have added the pink (red) band which refers to the axion star masses within the credible region at the $95\% (68\%)$ confidence to explain the single microlensing event in Subaru HSC data~\cite{Sugiyama:2021xqg}.}
\label{Q2}
\end{figure}
The conditions to be satisfied in the kinetic regime correspond to
\begin{align}
    (m_{a} v_{a}) \times (R_{\text{halo}}) \gg 1\,,\label{con1} \\
    (m_{a} v_{a}^2)\times (\tau_{\text{gr}}) \gg 1\,,\label{con2}
\end{align}
where $\tau_{\text{gr}} \sim m_{a}^3 v_{a}^6 \rho_{\text{halo}}^{-2}$ is the axion star condensation timescale.  The masses for the axion and the central PBH in minihaloes are related to each other so that conditions in Eqs.~(\ref{con1}) and (\ref{con2}) are satisfied at certain redshift. 

After axion stars are nucleated, they continue capturing axions from minihalo environment until reach saturation~\cite{Levkov:2018kau, PhysRevD.100.063528}. The saturated mass at redshift $z_{\star}$ shows the same relation with the minihalo mass $M_{\text{halo}}$ as those from solitonic core formation in fuzzy dark matter halos~\cite{Schive:2014hza} or axion clump nucleation in miniclusters~\cite{PhysRevD.100.063528}: $M_{\star}\propto \textcolor{black}{(1+z_{\star})^{1/2}}M_{\text{halo}}^{1/3}$~\cite{Hertzberg:2020hsz}.  
 
Take a flat $\Lambda$CDM cosmology and use values based on Planck TT, TE, EE+lowE+lensing+BAO at the $68\%$ confidence levels in Ref.~\cite{Planck:2018vyg}.
By simultaneously \textcolor{black}{satisfying} Eqs.~(\ref{con1}) and (\ref{con2}) at given initial redshift and calculating the corresponding condensation timescale, we show in Fig.~\ref{Q2} the parameter space $(m_{a}, M_{\star})$ (blue shaded region) associated with axion stars nucleation in minihalos before the time of first galaxies formation ($z_{\star}\gtrsim 30$). We have taken  $m_{a} = 2\times 10^{-11}\text{eV}$ as the lower bound for the axion mass considering the QCD  axion excluded region from stellar black hole spin measurements~\cite{Arvanitaki:2014wva}. In addition, we have added a pink (red) shaded band referring to the axion star masses within the credible region at the $95\% (68\%)$ confidence to explain the Subaru HSC observation~\cite{Sugiyama:2021xqg}. We see that the maximum axion star mass which can be nucleated is about $10^{-7}M_{\odot}$.

\textcolor{black}{Generally speaking, magnification of the light from source stars originated by extended lens is suppressed for finite size effects when the radius of the compact object is much larger than the corresponding Einstein radius. The microlensing efficiency of extended sources in comparison to the pointlike case is usually measured in terms of the  threshold impact parameter $u_{1.34}$, which refers to the impact parameter
for a lens such that all smaller impact parameters produce a magnification above the threshold~\footnote{\textcolor{black}{The threshold for the magnification adopted by HSC survey is 1.34 and it corresponds to the outer ray passing the lens at a radius of $1.618\times R_E$, e.g., the golden ratio times the Einstein radius, for the point-lens case.}}. By numerically solving the Schr$\ddot{\text{o}}$dinger-Poisson equation in the absence of self-interactions, the threshold impact parameter in terms of the ratio  $R_{\star}^{90}/R_E$ for ground state configurations is calculated in Ref.~\cite{Croon:2020wpr} (see Fig. 4 there).   
For $R^{90}_{\star}/R_{E} \gtrsim 5$, we have $u_{1.34}=0$.}~\footnote{\textcolor{black}{A similar result is found for the case in which the axion self-interaction is included, as shown  Fig. 3 (top panel) in Ref.~\cite{Fujikura:2021omw}. There an exponential-linear ansatz is used to approximate axion clump radial profiles. The shutoff of the magnification occurs at $R_E/R \lesssim 0.6$. Since $R = 3.610 R^{90}_{\star}$ for the particular ansatz~\cite{Schiappacasse:2017ham}, we have $R^{90}_{\star}/R_{E} \gtrsim 6$.}} 

\textcolor{black}{We may estimate the finite size lens effect of the axion clumps nucleated in axion minihalos around PBHs by using 
Eqs.~(\ref{ER}) and (\ref{m-r}) and the mass regime of axion star masses of our interest, $4\times 10^{-10}M_{\odot}\lesssim M_{\star}\lesssim 2\times10^{-7}$, to obtain $35 \lesssim R_{\star}/R_{E} \lesssim 80$.} Thus, the single microlensing event reported in the Subaru HSC observation cannot be linked to these astrophysical objects.

\section{Discussion and Conclusion}

We have shown that the known \textcolor{black}{mechanisms} for QCD axion clump formation in the pre- and postinflationary PQ symmetry-breaking scenario are not able to explain the microlensing event reported by the Subaru HSC observation~\cite{Niikura:2017zjd}. 

In the postinflationary scenario, the needed mass for axion stars is larger than the allowed maximum mass for a stable configuration. Since the axion clump masses run with the axion mass as $M_{\star}\propto m_{a}^{-2}$ and the QCD axion mass is strictly bounded from below due to the axion decay of topological defects, the typical axion star masses are $M_{\star} \lesssim 10^{-13} M_{\odot}$~\footnote{\textcolor{black}{In non-standard cosmologies, if the reheating temperature is sufficiently low and there is significant entropy production, the axion mass may reach smaller values than those in the standard case  ~\cite{Kawasaki:1995vt, Visinelli:2009kt}.  Including axion string decays and taking $N_{\text{DW}}=1$, the} \textcolor{black}{highest allowed value for the PQ scale reported in Ref.~\cite{Visinelli:2009kt} is $F_a\simeq 8.6\times10^{13}\text{GeV}$, when the misalignment mechanism gives the dominant contribution to the axion abundance. However, if higher axionic string contributions are considered, the found value is $F_a\simeq1.4\times10^{13}\text{GeV}$. While for the later case the maximum axion star mass is marginally within the HSC credible region shown in Fig. 1, for the more relaxed scenario we have $M_{\star}\lesssim 2\times 10^{-7}M_{\odot}$. The maximum mass reached by axion stars heavily depends on the model used to describe the axionic string evolution and the energy spectrum of emitted axions.} }. \textcolor{black}{Such mass} values are several orders of magnitude smaller than the typical mass needed for microlensing.  Thus, formation of axion stars through gravitational condensation in the early Universe via thermalization or axion  stars  nucleation  in  the  dense  core  of  axion  miniclusters cannot explain HSC data.

In the preinflationary scenario, the nucleation of axion clumps within axion self-similar minihalos
around PBHs produce axion stars with typical masses $4\times 10^{-10}M_{\odot}\lesssim M_{\star}\lesssim \textcolor{black}{2\times10^{-7}} M_{\odot}$, which is within the lightest part of the credible region associated with HSC data. However, such compact objects are related to light axion masses as $2\times 10^{-11}\,\text{eV} \lesssim m_{a} \lesssim 10^{-9}\,\text{eV}$. Since the axion star radius runs with the axion mass for a fixed axion star mass as $R_{\star} \propto m_{a}^{-2}$, the ratio between the axion star and Einstein radii is \textcolor{black}{$R_{\star}/R_E\sim \mathcal{O}(10)$}. The sizable finite lens size effect will \textcolor{black}{completely} suppress the magnification of the source star light as discussed in Refs.~\cite{Croon:2020ouk, Fujikura:2021omw}. 

Apart from the previous formation mechanisms for axion stars, there is still the possibility that
axion clumps are formed in the preinflationary scenario via thermalization and subsequent gravitational condensation, as they do  in the postinflationary scenario. Even though the axion field fluctuations are initially much smaller than those when the PQ symmetry is broken after inflation,
they could still growth via perturbation theory during the matter-dominated era and form a type of BEC in the late Universe. Such a complex analysis is beyond of the scope of this article, but we leave this inquiry for future work.

We point out that PBHs produced, for example, from the collapse of inflationary density perturbations in the early Universe,  are a plausible explanation for the single microlensing candidate reported by the Subaru HSC observation. 

\textcolor{black}{Axion miniclusters themselves have been proposed as sources for microlensing and first observational constraints over their abundance were calculated in Ref.~\cite{Fairbairn:2017sil} based on Subaru HSC~\cite{Niikura:2017zjd} and EROS~\cite{EROS-2:2006ryy} data. A remaining and interesting task is to determine and study the credible regions at different confidence levels to explain the associated microlensing candidates reported in such observations.} 

Last, as we mentioned previously in footnote 3, axionlike particle stars could also explain the microlensing candidate detected in the Subaru HSC survey because they can easily reach the required star mass range.   

\section{Acknowledgments}
This work was supported by the Academy of Finland Grant No. 318319.
E.D.S. thanks Kohei Fujikura for discussion.
 T.T.Y. is supported in part by the China Grant for Talent Scientiﬁc
Start-Up Project and the JSPS Grant-in-Aid for Scientiﬁc Research Grants No.
16H02176, No. 17H02878, and No. 19H05810 and by World Premier
International Research Center Initiative (WPI Initiative), MEXT, Japan.\\
$^\dagger$\href{mailto:edschiap@uc.cl}{edschiap@uc.cl}\\
$\ddagger$\href{mailto:tsutomu.tyanagida@sjtu.edu.cn}{tsutomu.tyanagida@sjtu.edu.cn}

\bibliography{main} 

\end{document}